\documentclass{aa}
\usepackage{graphicx}
\usepackage{amsmath}
\usepackage[round]{natbib}

\newcommand{\kms}{km s$^{-1}$}
\newcommand{\vsi}{\ensuremath{v \sin i}}
\newcommand{\bz}{\ensuremath{\langle B_z \rangle}}
\newcommand{\bs}{\ensuremath{\langle B \rangle}}
\newcommand{\bbz}{\ensuremath{\langle B_z^2 \rangle}}
\newcommand{\bbs}{\ensuremath{\langle B^2 \rangle}}
\newcommand{\gtsimeq}{\raisebox{-0.6ex}{$\,\stackrel
       {\raisebox{-.2ex}{$\textstyle >$}}{\sim}\,$}}
\begin{document}
\titlerunning{Measuring the surface magnetic fields of magnetic stars}
\title{Measuring the surface magnetic fields of magnetic stars with unresolved Zeeman splitting\thanks{Based on observations made with the European Southern Observatory telescopes under ESO programmes 60.A-9036(A), 68.D-0254(A), 69.D-0378(A), 074.D-0392(A), 075.C-0234(A), 075.C-0234(B), 076.C-0073(A), 076.D-0169(A), 077.D-0477(A), 079.D-0118(A), 079.C-0170(A), 079.C-0170(B), 080.C-0032(A), 080.C-0032(B), 081.C-0034(A), 081.C-0034(B), 082.C-0218(A), 082.C-0308(A), 089.C-0207(A), 089.D-0383(A), 266.D-5655(A), obtained from the ESO/ST-ECF Science Archive Facility. Based on observations obtained at the Canada-France-Hawaii Telescope (CFHT) which is operated by the National Research Council of Canada, the Institut National des Science de l'Univers of the Centre National de la Recherche Scientifique of France, and the University of Hawaii. Based in part on observations available on the ELODIE archive. }}
\author{J. D. Bailey\inst{1,2}}
\institute{Max-Planck-Institut f\"{u}r extraterrestrische Physik,
  Giessenbachstrasse 1, 85748, Garching,
  Germany\\ \email{jeffbailey@mpe.mpg.de}\label{inst1}
\and School of Physics and Astronomy, University of Leeds,
  Leeds, United Kingdom, LS2 9JT\label{inst2}}

\date{Received /
       Accepted }

\abstract {}
{High-dispersion, archival spectra of magnetic Ap stars
  with resolved Zeeman components in Stokes $I$ are used to derive a simple
  relation that can be utilised to estimate the mean surface field
  strengths of stars with \vsi\ \gtsimeq\ 10~\kms.}
{For each star, the
mean surface field, as measured from the observed splitting in Fe~{\sc
  ii} at 6149~\AA, is compared to the differential broadening of spectral lines with
large and small Land\'e factors in order to produce a relation to
estimate the field strengths of magnetic stars with unresolved Zeeman patterns.}
{The method is shown to be reliable for rotational velocities up to
  about 50~\kms\ for field strengths down to about
  5~kG. 
}
{These results should allow for better
constraints to be placed on the mean surface magnetic fields of Ap
stars where Zeeman patterns are unresolved.}

\maketitle

\section{Introduction}
The detections of magnetic fields in stars of the upper main sequence
rely on observing the Zeeman effect in spectral lines. This is
generally done by either observing polarisation or directly measuring Zeeman
splitting in stellar spectral lines. The
former furnishes the line-of-sight magnetic field, \bz, whereas the
latter gives the surface magnetic field, \bs. 

The mean line-of-sight (or longitudinal) magnetic field is a line
intensity weighted average over the visible hemisphere of a star of the magnetic field component directed
along the line-of-sight. It is obtained
from measuring the separation between the positions of the spectral
line profiles in left and right circularly polarised light. Observed
variations in \bz\ for a given star provide
constraints for the magnetic field structure (that is assumed to be symmetric about an
axis that is inclined to the rotation axis) such as the mean and
maximum/minimum \bz\ values.  However, \citet{Mathys2004} points out
that characterising the actual stellar magnetic field from the
longitudinal field alone is not possible due to the strong dependence
on the geometry of the observation. 

The mean surface magnetic field
strength can be measured from resolved Zeeman splitting in a spectral
line, providing a line intensity weighted average over the visible
stellar hemisphere of the modulus of the magnetic field. Zeeman splitting is difficult
to measure in spectral lines because even small rotational velocities can mask this signature. However, if the
rotational velocity broadening is sufficiently small compared to the magnetic
field splitting (usually the field strength in kG should be larger than
the rotational velocity in \kms), it is possible to measure \bs\ directly from observed Zeeman patterns in
a stellar spectrum \citep[see][]{LM2000}. This provides further
constraints on the magnetic field configuration (maximum and minimum \bs\
values) allowing for a more detailed picture of the overall field
structure \citep[see][]{Landstreet2009}. 

The magnetic chemically peculiar A-- type
stars (Ap) host the largest magnetic fields of stars on the main
sequence, with \bz\ ranging in strength from hundreds to thousands of
Gauss. Generally, sinusoidal variations in \bz\ are observed over
the rotational period of the star. The simplest model to describe these variations is
an oblique rotator that consists of the angles $i$ and $\beta$,
describing the inclinations of the line-of-sight and magnetic field
axes to the rotation axis, respectively \citep[see][for a clear
explanation of the rigid rotator model]{Landstreet1970}. Therefore, to completely specify the magnetic field
geometry requires knowledge of three parameters $i$, $\beta$ and the
polar field strength $B_{\rm d}$. As such, variations in \bz\
alone cannot supply sufficient constraints to specify a unique
geometry; however, with the addition of \bs, not only can we specify
all three parameters, but also additional field components such as a
magnetic quadrupole aligned with a magnetic
dipole \citep[see][for a more detailed
description]{Landstreet2009}. Thus, \bs\ is integral in providing the overall field
structure, but most often times Zeeman patterns are undetectable
because the star rotates too quickly. In such cases, stars are
diagnosed as being magnetic only from measurements of \bz. 

\citet{Mathys1995} combats the problem of unresolved Zeeman patterns
in magnetic stars by introducing the mean quadratic field, $B_{\rm mq}$. The mean
quadratic field is derived from the measure of the second order
moments of the Stokes $I$ parameter. It characterises the widths of
spectral lines about their centres due to magnetic broadening. In
particular, the mean quadratic field is the square root of the sum of the squares
of the mean surface magnetic field and the line-of-sight magnetic
field: $B_{\rm mq} = (\bbs + \bbz)^{0.5}$. The lower limit to the success of this method was found to be about 5~kG.  

\citet{Preston1971} suggests a simpler method in characterising the
mean surface fields of rotationally broadened stars, which also
exploits the fact that magnetic fields can broaden spectral lines such
that the broadening is significant compared to the dominant
rotational broadening. Preston measured the mean surface fields for
seven stars with resolved Zeeman patterns and for these same stars
compared the widths of spectral lines that are strongly affected by
the local magnetic field to ones that are not, finding a linear
correlation between spectral line widths and magnetic field strength. Recently, this relationship was used by \citet{Baileyetal2012} to constrain \bs\ in a star (HD~133880) with \vsi\ larger than 100~\kms.

This paper aims to improve upon the relation of \citet{Preston1971}
with the goal of extending this work to be used on rapidly rotating
magnetic Ap stars to better constrain their magnetic fields without
having to rely solely on line-of-sight magnetic measurements from
circular polarisation. This is achieved by utilising an extensive
collection of high dispersion spectra of magnetic stars with resolved
Zeeman splitting from the ESO and CFHT archives: a total of 18 stars
and 102 spectra. This study also benefits from the much higher
signal-to-noise ratio ($SNR$) and increased linearity of today's
instruments. The following section discusses the
observations. Sections 3 and 4 describe the methods and results,
respectively. Section 5 tests the method by comparing to synthetic spectra and Section 6 outlines the conclusions from this work.

\section{Observations}
\begin{table}
\centering
\caption{The star designations, instrument (with the number of spectra indicated in parentheses), spectral resolution, and spectral range are listed for the stars used in this study.}
\begin{tabular}{lrrr}
\hline\hline
Star & Instrument & R & $\lambda$ (\AA) \\
\hline
HD~2453 & ELODIE (1) & $42\,000$ &  $3900-6800$\\
               & UVES (1)& $80\,000$ & $3282-4563$ \\
               &           &$110\,000$ & $5708-9463$\\
HD~12288 & ELODIE (1)& $42\,000$ &  $3900-6800$\\
HD~51684 & UVES (2) & $80\,000$ & $3282-4563$\\
                  &              &$110\,000$ & $4727-6835$\\
HD~65339 & ESPaDOnS (1) & $65\,000$ & $3690-10481$\\
HD~93507 & HARPS (12) & $115\,000$ & $3780-6910$ \\
HD~94660 & ESPaDOnS (1) &$65\,000$ & $3690-10481$\\
                  & HARPS (2)      & $115\,000$& $3780-6910$ \\
HD~116114 & HARPS (14) &  $115\,000$& $3780-6910$ \\
HD~116458 & HARPS (10) & $115\,000$& $3780-6910$ \\
HD~126515 & UVES (1) & $80\,000$ & $3731-4999$\\
                    &               &$110\,000$& $4726-6835$\\
                    & HARPS (10) & $115\,000$& $3780-6910$ \\
HD~137909 & ESPaDOnS (4) & $65\,000$ & $3690-10481$\\
                    & HARPS (1)      & $115\,000$& $3780-6910$ \\
HD~142070 & HARPS (6) &  $115\,000$& $3780-6910$ \\
HD~144897 & HARPS (5) & $115\,000$& $3780-6910$ \\
HD~166473 & UVES (1) & $80\,000$ & $3731-4999$\\
                    &               &$110\,000$& $4726-6835$\\
                    & HARPS (7) & $115\,000$& $3780-6910$ \\
HD~187474 & UVES (1) & $80\,000$ & $3731-4999$\\
                    &               & $110\,000$& $4727-6835$\\
HD~188041 & HARPS (2) & $115\,000$& $3780-6910$ \\
HD~192678 & ELODIE (1) & $42\,000$ &  $3900-6800$\\
HD~208217 & HARPS (9) & $115\,000$& $3780-6910$ \\
HD~318107 & ESPaDOnS (2) &$65\,000$ & $3690-10481$\\
                    & HARPS (7)       &$115\,000$& $3780-6910$ \\
\hline
\label{stars}
\end{tabular}
\end{table}
The majority of data used for this study were obtained from the European Southern Observatory (ESO)
and Canada-France-Hawaii Telescope (CFHT) archives (using the UVES, HARPS and ESPaDOnS instruments), with
selected spectra also taken from the ELODIE archive, for a total
of 102 high resolution spectra for 18 magnetic stars with resolved
Zeeman patterns. Table~\ref{stars} lists the stars used for this study. Below I describe each instrument in brief.

ELODIE was a spectrograph located at Observatoire de Haute Provence. It covered a spectral range from about $3900$ to $6800$~\AA\ with a resolving power of $R \simeq 42\,000$. All the ELODIE spectra utilised in this study have $SNR$ of about 100.

ESPaDOnS, a cross-dispersed echelle spectropolarimeter, is located at the CFHT. The instrument covers a spectral range from $3690$ to $10\,481$~\AA\ with a resolving power of $R \simeq 65\,000$ (in polarimetric mode). The available spectra have $SNR$ above about 200.

Located at the ESO La Silla 3.6 m telescope, HARPS is a cross-dispersed echelle spectrograph covering a spectral range of 3780 - 6910~\AA\ with $R \simeq 115\,000$.  The $SNR$ of the available spectra were also of order 200. 

UVES is a cross-dispersed optical spectrograph that is located at ESO's Paranal Observatory. It has both a blue and red arm that offers different spectral resolutions and range coverage. The blue arm, with $R \simeq 80\,000$, covers a spectral range from about $3100$ to $4900$~\AA, whereas the red arm ($R \simeq 110\,000$) can range from about $4800$ to $10\,200$~\AA. The $SNR$ of the available spectra were above about 300.

\section{Measurements}
I endeavour to obtain an estimate for \bs\ from comparing the widths
of two spectral lines that have large and small Land\'e factors $z$
(i.e. lines that are strongly affected by the local magnetic field to
ones that are not). To achieve this, I adapt the measurement technique
of \citet{Preston1971} which is summarised below.
\subsection{The $K$--parameter}
If one considers the $\sigma$--$\sigma$ separation of Zeeman split components, it is
straightforward to show that the width of a line broadened by the magnetic field
can be written as $const \times \lambda^{2}z$. \citet{Preston1971}
defines this constant as the parameter $K$, which is proportional to
the strength of the magnetic field, \bs. Using the same notation as
\citet{Preston1971}, if it is further assumed that
the instrumental ($w_{I}$) and magnetic broadening profiles can be added in
quadrature then
\begin{subequations}
\begin{align}
w_{L}^{2} = w_{I}^{2} + K^{2}\left<\lambda^{4}z^{2}\right>_{L}, \label{eq:a}\\
w_{S}^{2} = w_{I}^{2} + K^{2}\left<\lambda^{4}z^{2}\right>_{S}, \label{eq:b}
\end{align}
\end{subequations}
where $w_{L}$ and $w_{S}$ are the widths for lines with large and
small $z$ values, respectively. Combining Equations~1a and 1b, the $K$
parameter can be written as
\begin{equation}
K = \left(\frac{w_{L}^{2} - w_{S}^{2}}{\left<\lambda^{4} z^{2}\right>_{L} - \left<\lambda^{4} z^{2}\right>_{S}}\right)^{1/2}.
\end{equation}
Note that for this relation, both the widths and wavelengths are
measured in centimetres meaning that $K$ is the wavenumber (i.e. units of
cm$^{-1}$), which are the same units used by \citet{Preston1971}.

\subsection{Technique}
For each star listed in Table~\ref{stars}, the mean magnetic surface
fields were measured from the observed splitting in the Fe~{\sc ii}
6149 line. In addition to this, for each star the widths of up to five
spectral lines in Table~\ref{lines} were measured. To obtain accurate
and consistent measurements, I had to consider that the lines with
large $z$ values could have multiple Zeeman components which would
make fitting a simple Gaussian infeasible. Therefore, for each line
the full width at half--depth (FWHD) was measured. The widths of each line
were then used to measure the $K$--parameter from Equation~2 with the
necessary atomic data collected from the Vienna Atomic Line Database
\citep[VALD;][]{vald1,vald2,vald3,vald4}. For each spectrum, an average
$K$ value was found from all possible combinations of lines with large and
small $z$ values with an estimated uncertainty that corresponds
  to the largest spread in $K$ values from different sets of lines. The uncertainties in \bs\ measured from Fe~{\sc ii} 6149
were estimated by considering the scatter about the mean fit of multiple
individual measurements for each spectrum. Table~\ref{results} lists these average $K$ values
and \bs\ from Fe~{\sc ii} 6149, with associated uncertainties, for
each star. For stars (or multiple spectra for the same star) which
  exhibit little variation between \bs\ measurements, only mean
  values for \bs\ and $K$ are recorded in Table~\ref{results}. These
  are denoted by an asterisk.

\begin{center}
\begin{table}
\centering
\caption{Fe~{\sc ii} lines used to measure widths. Listed are the rest wavelengths and Land\'e $z$ factors.}
\begin{tabular}{rr|rr}
\hline\hline
 & & & \\
\multicolumn{2}{c|}{Large $z$} & \multicolumn{2}{c}{Small $z$}\\
\hline
$\lambda$ (\AA) & $z$ & $\lambda$ (\AA) & $z$ \\
\hline
4271.404 & 1.33 & 4491.400 & 0.40\\
4303.176 & 1.43 & 4508.280 & 0.50\\
4520.220 & 1.34 & & \\
\hline
\label{lines}
\end{tabular}
\end{table}
\end{center}

\section{Results}
\begin{center}
\begin{figure}
\centering
\includegraphics[angle=-90, width=0.5\textwidth]{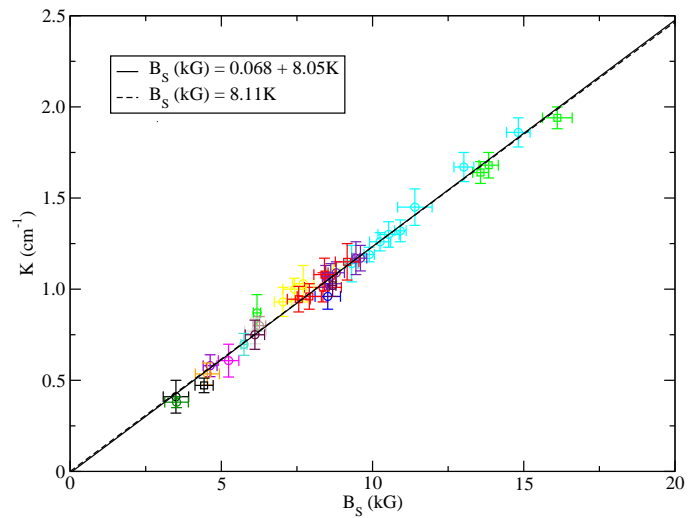}
\caption{Shown are the measured $K$-parameters versus the surface field
  strengths (as measured from observed splitting in Fe II 6149) from
  the values in Table~\ref{results}. Each
  colour and/or shape indicates measurements from an individual
  star. The solid line is the best-fit line of all the data and the
  dotted black line is the best-fit when forcing the y-intercept (i.e. K)
  through the origin. }
\label{K-Bs-relation}
\end{figure}
\end{center}
\begin{table}
\centering
\caption{Shown are the $K$ value, \bs\ measurements and estimated
  surface field strengths from Eqn~3, with
  associated uncertainties, for each star. For stars with spectra that exhibit
  little \bs\ variability, an average value of multiple spectra is shown. }
\begin{tabular}{lllr}
\hline\hline
Star & $K$ (cm$^{-1}$) & \multicolumn{2}{c}{\bs\ (kG)} \\
       &                       &Fe~{\sc ii}~6149 & Eqn.~3\\ 
\hline
HD~2453 & 0.41 $\pm$ 0.09* & 3.50 $\pm$ 0.42* & 3.33 $\pm$ 0.73 \\
HD~12288 & 1.01 $\pm$ 0.08 & 8.36 $\pm$ 0.61 & 8.19 $\pm$ 0.65 \\
HD~51684 & 0.87 $\pm$ 0.10* & 6.18 $\pm$ 0.12* & 7.06 $\pm$ 0.81 \\
HD~65339 & 0.96 $\pm$ 0.07 & 8.52 $\pm$ 0.42 & 7.79 $\pm$ 0.57\\
HD~93507 & 0.93 $\pm$ 0.08* & 7.04 $\pm$ 0.29* & 7.54 $\pm$ 0.65 \\ 
                  & 1.00 $\pm$ 0.06* & 7.42 $\pm$ 0.39* & 8.11 $\pm$ 0.49\\
                  & 1.03 $\pm$ 0.10* & 7.71 $\pm$ 0.35* & 8.35 $\pm$ 0.81\\
                  & 1.09 $\pm$ 0.05* & 8.70 $\pm$ 0.22* & 8.84 $\pm$ 0.41\\ 

HD~94660 & 0.80 $\pm$ 0.05* & 6.26 $\pm$ 0.21* & 6.49 $\pm$ 0.41\\
HD~116114 & 0.77 $\pm$ 0.07* & 6.13 $\pm$ 0.36* & 6.24 $\pm$ 0.57\\
HD~116458 & 0.58 $\pm$ 0.06* & 4.63 $\pm$ 0.23* & 4.70 $\pm$ 0.49\\
HD~126515 & 1.14 $\pm$ 0.10* & 9.31 $\pm$ 0.21* & 9.25 $\pm$ 0.81 \\
                                  & 1.19 $\pm$ 0.04 & 9.89 $\pm$ 0.17 & 9.65 $\pm$ 0.32 \\
                                  & 1.26 $\pm$ 0.05 & 10.25 $\pm$ 0.35 & 10.22 $\pm$ 0.41 \\
                                  & 1.30 $\pm$ 0.07 & 10.53 $\pm$ 0.35 & 10.54 $\pm$ 0.57 \\
                                  & 1.32 $\pm$ 0.06 & 10.92 $\pm$ 0.19 & 10.71 $\pm$ 0.49 \\
                                  & 1.45 $\pm$ 0.10 & 11.40 $\pm$ 0.57 & 11.76 $\pm$ 0.81\\
                                  & 1.67 $\pm$ 0.08 &13.02 $\pm$ 0.33 & 13.54 $\pm$ 0.65 \\
                                  & 1.86 $\pm$ 0.08* & 14.82 $\pm$ 0.39* & 15.08 $\pm$ 0.65 \\

HD~137909 & 0.61 $\pm$ 0.09* &  5.24 $\pm$ 0.34* & 4.93 $\pm$ 0.73 \\
HD~142070 & 0.54 $\pm$ 0.06* &  4.54 $\pm$ 0.40* & 4.34 $\pm$ 0.49\\
HD~144897 &1.07 $\pm$ 0.06 & 8.45 $\pm$ 0.11 & 8.68 $\pm$ 0.49 \\
                    & 1.07 $\pm$ 0.07 & 8.62 $\pm$ 0.16 & 8.68 $\pm$ 0.57 \\
                    & 1.09 $\pm$ 0.06 & 8.80 $\pm$ 0.27 & 8.84 $\pm$ 0.49 \\
                    & 1.17 $\pm$ 0.09 & 9.45 $\pm$ 0.13 & 9.49 $\pm$ 0.73 \\
                    & 1.17 $\pm$ 0.07 & 9.60 $\pm$ 0.21 & 9.49 $\pm$ 0.57\\

HD~166473 & 0.75 $\pm$ 0.08* & 6.11 $\pm$ 0.32* & 6.08 $\pm$ 0.65\\
                                   & 1.02 $\pm$ 0.06 & 8.68 $\pm$ 0.11 & 8.27 $\pm$ 0.49 \\
HD~187474 & 0.70 $\pm$ 0.06 & 5.75 $\pm$ 0.10 & 5.65 $\pm$ 0.49\\
HD~188041 & 0.38 $\pm$ 0.03* & 3.52 $\pm$ 0.39* & 3.08 $\pm$ 0.24\\
HD~192678 & 0.47 $\pm$ 0.04 & 4.43 $\pm$ 0.30  & 3.83 $\pm$ 0.33\\
HD~208217 & 0.94 $\pm$ 0.07 & 7.56 $\pm$ 0.38 & 7.66 $\pm$ 0.57\\
                    & 0.96 $\pm$ 0.07 & 7.91 $\pm$ 0.40 & 7.79 $\pm$ 0.57\\
                    & 1.08 $\pm$ 0.09* & 8.41 $\pm$ 0.35* & 8.76 $\pm$ 0.73\\ 
                    & 1.15 $\pm$ 0.10* & 9.16 $\pm$ 0.39* & 9.33 $\pm$ 0.81\\

HD~318107 & 1.64 $\pm$ 0.06 & 13.57 $\pm$ 0.26  & 13.30 $\pm$ 0.48 \\
                                  & 1.68 $\pm$ 0.07* & 13.84 $\pm$ 0.32* & 13.62 $\pm$ 0.57\\
                                  & 1.94 $\pm$ 0.06 & 16.11 $\pm$ 0.49&  15.73 $\pm$ 0.49 \\ 
\hline
\multicolumn{4}{p{0.4\textwidth}}{{\sc Notes.} (*) Denotes an average
  of multiple spectra.} \\
\label{results}
\end{tabular}
\end{table}
Figure~\ref{K-Bs-relation} plots the relation between $K$ and \bs, showing a clear linear relation. The best-fit line (solid black) of all
the data suggests that $\bs\ {\rm (kG)} =  0.068 + 8.05K$. However, for
a star without a magnetic field, in which $K$ is zero, the measured
field from the relation should also be zero. Therefore, also shown is
the linear fit with the y-intercept forced through the origin (black
dashed line). Both lines fit the data
equally well; therefore, since the fitted relation should pass
  through the origin, the latter relation with a zero constant term is
adopted with
\begin{equation}
\bs\ {\rm (kG)} = 8.11K.
\end{equation}
This equation is used to calculate the surface magnetic field strength
in Table~\ref{results}.

\subsection{Effects of rotational broadening}
The goal of this work is to measure $K$ values for stars with larger
\vsi\ (\gtsimeq\ 20~\kms) in order to estimate
\bs. Therefore, it is important to ascertain how feasible it is to
measure magnetic broadening in spectral lines as rotational broadening
increases. To investigate this, it is necessary to compare spectra
with and without the effects of a magnetic field. {\sc zeeman} is a {\sc
  fortran} program that synthesises stellar spectra and is capable of including
the effects of a magnetic field
\citep[detailed descriptions of {\sc zeeman} are provided
by][]{Landstreet1988,Wadeetal2001}. 

\begin{center}
\begin{figure}
\centering
\includegraphics[angle=-90, trim = 0in 0in 0in 4.5in, clip, width=0.55\textwidth]{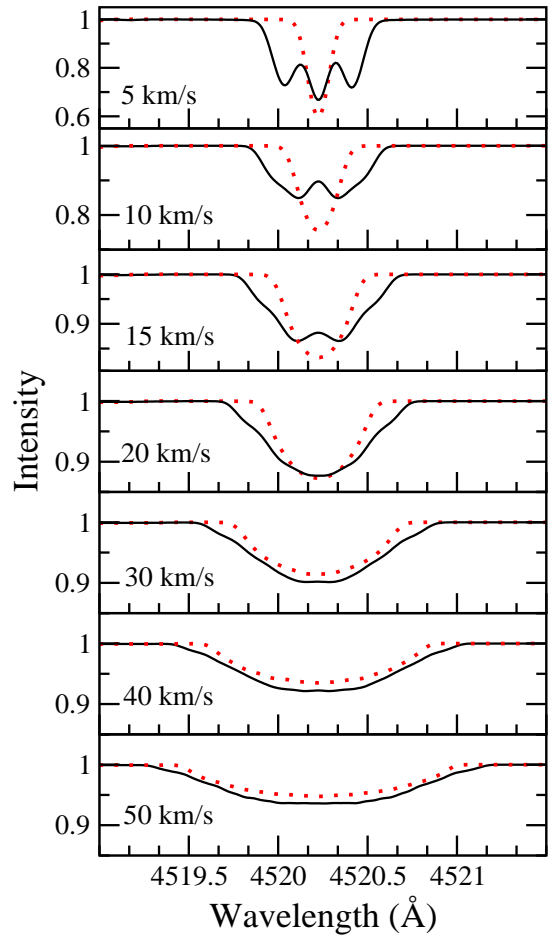}
\caption{Shown are comparisons between the Fe~{\sc ii} 4520 line for a
15~kG field (black lines) and no magnetic field (red dots) for
increasing values of \vsi.}
\label{zeeman-vsi}
\end{figure}
\end{center}

{\sc zeeman} was used to create
pairs of synthetic spectra for Fe~{\sc ii} 4520 (with and without the
presence of a magnetic field) for incrementally increasing values of
\vsi. For this comparison, a field strength of 15~kG was used. The results are shown in Figure~\ref{zeeman-vsi}. One can see how the Zeeman
splitting in the Fe~{\sc ii} 4520 line slowly disappears as \vsi\ increases, but magnetic
broadening still remains. Above about 50~\kms, the magnetic broadening
is only marginally larger than the rotational
broadening. Table~\ref{hwidth} lists the FWHD for the lines in
Figure~\ref{zeeman-vsi}, with typical uncertainties of order
0.005~\AA, as determined from the scatter of multiple measurements. Even for values of \vsi\ of about
50~\kms, there is still a measurable difference between the
rotationally broadened and magnetically and rotationally broadened spectral line of
Fe~{\sc ii} at 4520~\AA; however, this difference is marginal,
suggesting that it may be difficult to obtain an accurate measure of
the field above about 50~\kms. Note that Fe~{\sc ii} 4520 has a
Land\'e factor $z = 1.34$ and that the choice of a different spectral line
with a larger $z$ value would increase the difference, perhaps
making it plausible to measure the field in stars with even larger
rotation rates.

\begin{table}
\centering
\caption{Measured full width at half depth (FWHD) for Fe~{\sc ii}~4520
at multiple values of \vsi\ for a field strength of 15~kG. Shown are the results without and with a
magnetic field and the difference between the two FWHD. The
measurements have typical uncertainties of order 0.005~\AA. }
\begin{tabular}{lrrr}
\hline\hline
 \vsi\ & \multicolumn{3}{c}{FWHD (\AA)} \\
 (\kms) & No $B$ & $B$ & $\Delta$width \\
 \hline
5   & 0.135 & 0.513 & 0.378\\
10 & 0.234 & 0.568 & 0.334\\
15 & 0.350 & 0.580 & 0.230\\
20 & 0.472 & 0.614 & 0.142\\
30 & 0.720 & 0.784 & 0.064\\
40 & 0.947 & 0.981 & 0.034\\
50 & 1.197 & 1.207 & 0.010\\
\hline
\label{hwidth}
\end{tabular}
\end{table}

\subsection{Artificial broadening of real spectra}
\begin{center}
\begin{figure}
\centering
\includegraphics[angle=-90, width=0.5\textwidth]{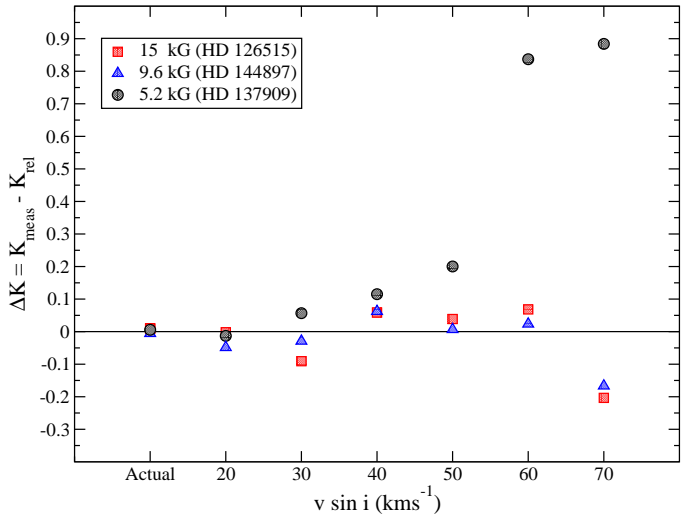}
\caption{Shown is the difference between the measured and predicted
  $K$-parameters for three different field strengths at various
  rotation rates. The actual measurement, when deriving the relation
  of Equation~2 is also shown. The black circles are for HD~137909
  with \bs\ $= 5.2$~kG, the blue triangles are for HD~144897 with \bs\
$=9.6$~kG and the red square are for HD~126515 with \bs\ $= 15$~kG.}
\label{K-Bs-cal}
\end{figure}
\end{center}
The preceding section suggests that reliably measuring magnetic
broadening in a spectral line may be difficult above a rotational
velocity of about 50~\kms. Therefore, in order to ascertain the reproducibility of the calculated $K$ values
at large \vsi, the spectra of three separate stars with field strengths around 5, 10
and 15~kG (HD~137909, HD~144897 and
HD~126515, respectively) were convolved with the rotational broadening
function \citep[e.g.][]{Unsold1955}. The spectra were broadened up to 70~\kms\ in
intervals of 10~\kms. For each spectrum, the $K$ values were measured
in the same manner as described in Section~3. Figure~\ref{K-Bs-cal} shows the
difference between the measured $K$ values and the $K$ value
predicted by Equation~2 plotted against \vsi\ for the rotationally
broadened spectra. Also shown is the actual
measurement obtained when deriving the relation.  

At both 10 and 15~kG, the $K$ value is consistently reproduced at all \vsi,
with a slightly larger disparity at 70~\kms\ of about 0.2~${\rm cm^{-1}}$. Above about
50~\kms, the $K$ value cannot be reliably measured for a field of
about 5~kG. Apparently the marginal difference in the FWHD at larger
\vsi\ shown in Figure~\ref{zeeman-vsi} and Table~\ref{hwidth} is sufficient to reproduce the measured
$K$ value from magnetic broadening for a field above about 10~kG up to
about 60 or 70~\kms. It is important to note that although
  Figure~3 suggests that larger magnetic fields follow the linear
  relationship of Equation~3 above \vsi\ of about 50~\kms, no a priori knowledge of
  the field strength will exist when a measurement is made for an
  actual star. Therefore, for the \vsi's considered, the method becomes
  unreliable above about 50~\kms\ for \bs\ less than or equal to about
  15~kG because it is impossible to know
  whether the measurement yields a reliable \bs\ value or not. 

\section{Comparison to synthetic spectra}
The reproducibility of the $K$ value at large \vsi's suggests that Equation~3 may be reliably used to estimate the
field strengths of magnetic stars that are fast rotators. In
Section~4.2, observed spectra were artificially broadened to determine
how accurately the $K$ parameter can be reproduced with increasing \vsi.
However, the question still remains of how accurate the relation is at measuring
the surface field of a star for which the field strength is known
precisely. To test this, I used the program {\sc zeeman} (see Sect. 4). In
addition to including the effects of magnetic fields, {\sc zeeman}
also allows as input the specification of a simple magnetic field geometry that is a
simple co-axial multipole expansion consisting of dipole,
quadrupole and octupole components with the angles between the
line-of-sight and rotation axis ($i$) and the magnetic field and
rotation axis ($\beta$) specified. As output, {\sc zeeman} also
provides the precise surface field strength. Therefore, it is possible
to compare the estimated field strength from Equation~3 to spectra for
which the exact field value is known.

Only Fe lines are used to measure $K$ and since
the magnetic stars in this study are most certainly Ap in nature, the
abundance of Fe was set to a value that is 0.5~dex above the solar
ratio. To test the effects of varying geometries, synthetic spectra
for \vsi\ 's up to 70~\kms\ were produced for simple dipolar field
strengths of 5, 10, 15 and 20~kG with $i =$ 0, 45 and 90
degrees. To test the applicability of this method of magnetic field determination
for increasing \vsi, a geometry where the magnetic field axis is
perpendicular to the rotation axis (i.e. $\beta = 90^{\circ}$) was favoured,
because the angle between the rotation and magnetic field axes are
statistically larger in faster rotators \citep{LM2000}. It is
recognised that the surface field measured depends not only on the
mean value \bs, but also, to some extent, on the organisation of the
field. However, the precision of this method is insufficient to detect
these types of changes. As such,
the phase dependence in measuring \bs\ with this method is not tested
and all computations are performed at the same phase,
ensuring that the line-of-sight lies in the plane defined by the
rotation and magnetic field axes.

\begin{center}
\begin{figure}
\centering
\includegraphics[angle=-90, width=0.55\textwidth]{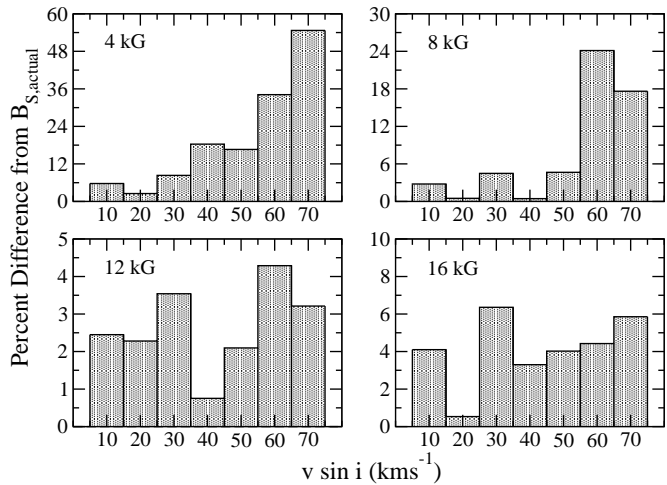}
\caption{Shown are the percent differences between the
measured \bs\ values from Equation~3 to the actual \bs\ value from the
synthetic spectra (as reported by {\sc zeeman}) plotted against \vsi. The approximate field strength is shown
in each panel. Note that the y-axis is different for each panel.}
\label{synth-compare}
\end{figure}
\end{center}

Figure~\ref{synth-compare} shows the percent difference between the
measured \bs\ values from Equation~3 to the actual \bs\ value from the
synthetic spectra (as reported by {\sc zeeman}) plotted against \vsi. The approximate field strength is shown
in each panel. Note that no significant differences were found between the three different geometries and therefore only the results
for the pole-on geometry are shown. For fields greater than
about 12~kG, the predicted field strength from Equation~3 accurately
predicts the surface field strength to within 6\% at all
\vsi\ considered. This is also true up to a rotation of 50~\kms\ for a field of
about 8~kG. Above this \vsi, the measured value is within 25\%  for an
8~kG field. At the lower field strength of 4~kG, the field is
predicted within 20\% at or below a rotation of 50~\kms\ and quickly
diverges at higher \vsi\ values. These results can be understood by
considering that at lower field values it becomes increasingly
difficult to discern between rotational and magnetic broadening in a
spectral line whereas for larger field values the magnetic broadening
is still significant compared to the rotational broadening (see Figure~\ref{zeeman-vsi}).

\section{Conclusions}
In this paper, a relation to estimate the surface magnetic field \bs\
for stars with large \vsi\ (above about 10~\kms) was derived. This was
done by extending the work of \citet{Preston1971} to include 102 total
spectra for 18
magnetic stars with resolved Zeeman splitting. For
each spectrum, the surface field strengths were measured from the
observed splitting in Fe~{\sc ii} 6149 and the widths of spectral
lines with large and small Land\'e factors were measured to determine
the $K$ parameter (a constant that is proportional to the magnetic
field strength). The derived relation is presented in Equation~3. 

The $K$ parameter was shown to be reliably reproduced at increasingly large \vsi\
when the spectra were rotationally broadened. For lower field values
(around 5~kG), the $K$ parameter is not accurately reproduced above
about 50~\kms. 

The relation of Equation~3 was also compared to synthetic spectra for
which the surface fields were known precisely. It was found that for
\bs\ greater than about 8~kG the field was accurately predicted
to within 6\% of the actual value. Below about 10~kG, the ability to
accurately measure the magnetic field above \vsi\ $\simeq 50$~\kms\
diminishes to within 25\% and is completely unreliable for surface
fields of order 8 and 4~kG, respectively. Below about 50~\kms, a 4~kG field can be
measured to within about 20\% with increased precision as \vsi\
decreases. However, since no knowledge of the surface field strength
exists prior to measuring the field from this technique, the method is
only useful up to \vsi\ of about 50~\kms\ and for magnetic field
strengths down to about 5~kG. 

These results are very encouraging and demonstrate that this simple method
can reliably estimate the surface field strengths of magnetic stars
with large \vsi\ (up to about 50~\kms). Thus, this technique will allow constraints to be placed
on the mean surface field variations of magnetic stars for which only
line-of-sight measurements were previously possible. 
\begin{acknowledgements}
JDB acknowledges support from the University of Leeds and thanks the
referee Gautier Mathys for helpful comments. JDB is also grateful to John D. Landstreet for many useful discussions and graciously reviewing the manuscript. Lastly, JDB thanks Drs Nicole Bailey and Andy Pon for many helpful hints with IDL. 
\end{acknowledgements}

\bibliographystyle{aa}
\bibliography{bs}
\end{document}